\documentclass[epj]{svjour}
\usepackage{graphics}
\def\la{\langle}
\def\ra{\rangle}
\def\beq{\begin{equation}}
\def\eeq{\end{equation}}
\def\be{\begin{eqnarray}}
\def\ee{\end{eqnarray}}

\def\k2av{\la k_T^2\ra}
\newcommand{\f}[2]{\frac{#1}{#2}}
\newcommand{\dd}{ {\textrm d}}
\begin{document}
\title{Jet Tomography in the Forward Direction at RHIC}
\author{G.G. Barnaf\"oldi\inst{1}\inst{2} 
\thanks{\emph{E-mail address:} bgergely@rmki.kfki.hu}%
\and P. L\'evai\inst{1} 
\and G. Papp\inst{3} \and G. Fai\inst{4}
}                     
%
%
\institute{
RMKI KFKI, P.O. Box 49, Budapest 1595, Hungary \and
Dept. of Physics of Complex Systems, E\"otv\"os University, 
P\'azm\'any P. 1/A, Budapest 1117, Hungary \and
Dept. for Theoretical Physics, E\"otv\"os University, 
P\'azm\'any P. 1/A, Budapest 1117, Hungary \and
Center for Nuclear Research, 
Kent State University, Kent, OH 44242, USA}
\date{Received: date / Revised version: date}
%
\abstract{
Hadron production at high-$p_T$ displays a
strong suppression pattern in a wide rapidity region 
in heavy ion collisions at RHIC energies.
This finding indicates the presence of strong final state effects for both
transversally and longitudinally traveling partons, namely 
induced energy loss. We have developed
a perturbative QCD based model to describe hadron production in
$pp$ collision, which can be combined with the
Glauber\,--\,Gribov model to describe hadron production in
heavy ion collisions. Investigating $AuAu$ and $CuCu$ collisions
at energy $\sqrt{s}=200$ $A$GeV at mid-rapidity, we find the
opacity of the strongly interacting hot matter to be proportional
to the participant nucleon number. Considering forward rapidities,
the suppression pattern indicates the formation of a longitudinally
contracted dense deconfined zone in central heavy ion collisions.
We determine parameters for the initial geometry from the existing data.}

\maketitle
\section{Introduction}
\label{intro}

One of the most interesting challenges in high-energy 
heavy ion collisions is to extract the properties of the 
produced hot dense medium. The discovery of strong pion 
suppression at RHIC energies in $AuAu$ 
collisions~\cite{PHENIX_QM01,PHENIX_QM05,STAR_QM05}
and its detailed analysis~\cite{Levai_QM01,GV02} 
was the first milestone along the way to determine 
the parameters of the central 
space-time region of heavy ion collision.
Applying the so called "jet-tomography" 
method~\cite{glv,zakhar,bdms,urs} one can "measure" the averaged density  
of the color charges in the deconfined region on the basis
of the induced energy loss of partons traversing the dense region.
Previously we have investigated the induced energy loss
for transversally propagating partons~\cite{bgg04}
analyzing mid-rapidity data at different centralities in $AuAu$ 
collisions~\cite{PHENIX03,PHENIX03a} and we have determined
the opacity parameter of the hot dense matter in the transverse direction. 
The next level is to perform tomography into the forward (or backward) 
directions and to extract integrated information about the 
$3$-dimensional structure of the color charge density.

In this paper we display forward tomography 
in $AuAu$ collisions at RHIC energy. 
Our aim is to understand the latest BRAHMS result~\cite{BRAHMS05}
obtained in $AuAu$ collisions, where the measured nuclear modification factor, 
$R_{AA}(p_T)$, displays similar values at different 
rapidities ($y=0$ and $y=3.2$). 
The explanation of this requires the understanding
of competing nuclear effects, namely
shadowing, multiscattering, induced energy loss,
and the qualitative consideration of the geometry of the dense
partonic region. On the other hand, these data offer a unique
possibility to extract information about the properties of the
dense matter in the longitudinal direction
in the very early stage of the heavy ion collisions.

Our survey will start with the $pp$ and the $dAu$ data as baseline.
At large forward rapidities the indication of increasing suppression 
has been seen in $dAu$ collisions in pion and charged
hadron production~\cite{BRAHMS,BRAHMSdau04}. The analysis of these
data helped us include the appropriate shadowing 
at high rapidities~\cite{Barnafqm04,bggqm05}. Thus we are able to
determine quantitatively the appropriate energy loss in $AuAu$ collisions,
assuming that all final state effects are connected to induced radiative
energy loss and the geometry of the dense matter.

We briefly introduce our perturbative QCD improved parton 
model~\cite{bgg04,Yi02} and the description of different nuclear effects.
The initial state nuclear shadowing and multiple scattering 
effects are able to reproduce the $\pi^0$ spectrum in 
$dAu$ collisions~\cite{Barnafqm04,dAu}.    
The final state radiative parton energy loss is modeled  
by the GLV-method~\cite{glv}. We extract the opacity of the dense matter 
in different centrality regions in $AuAu$ and $CuCu$ 
collisions at $\sqrt{s}=200$ $A$GeV at mid-rapidity.
On the basis of the extracted opacity values we perform the forward
tomography calculations and investigate existing $AuAu$ data
at large forward rapidities.

We discuss the obtained results and connect them to the longitudinal
geometry of the dense region. Our analysis is based on the
application of radiative energy loss of jets in the mid-rapidity and
forward rapidities. The role of other nuclear effects
(e.g. possible appearance of scattering energy loss for 
light quarks~\cite{Colum05,Colum06})
will be discussed in a forthcoming paper.


\section{Theoretical Description of the Model}
\label{sec:1}

The high-$p_T$ inclusive pion spectrum in a heavy ion 
collision can be calculated in a pQCD-improved parton model. 
Originally this was developed for nucleon-nucleon collision 
(basically $pp$), and extended by a Glauber-type collision geometry 
and initial state nuclear effects for nuc\-le\-us-nucleus, $AA'$ 
collisions as~\cite{Aversa89,Aur00,pgNLO}:
\begin{eqnarray}
\label{hadX}
&E_{\pi}& \f{\dd \sigma_{\pi}^{AA'}}{ \dd ^3p} = 
            \int \dd ^2b \,\, \dd ^2r \,\, t_A(r) 
            \,\, t_{A'}(|{\bf b} - {\bf r}|) \times \nonumber \\
& \times &  \f{1}{s} \,\, \sum_{abc} 
\int^{1-(1-v)/z_c}_{vw/z_c} \! \f{\dd \hat{v}}{\hat{v}(1-\hat{v})} \!
            \int^{1}_{vw/\hat{v}z_c} \f{ \dd \hat{w} }{\hat{w}} \!
            \int^1 {\dd z_c} \times \nonumber \\
& \times &  \int \!\! {\dd^2 {\bf k}_{Ta}} \!\! \int \!\! {\dd^2 {\bf k}_{Tb}}
            \,\, f_{a/A}(x_a,{\bf k}_{Ta},Q^2)
            \,\, f_{b/A'}(x_b,{\bf k}_{Tb},Q^2)  \nonumber \\
& \times &  \left[
            \f{\dd {\widehat \sigma}}{\dd \hat{v}} \delta (1-\hat{w})\, + \,
            \f{\alpha_s(Q_r)}{ \pi}  
            K_{ab,c}(\hat{s},\hat{v},\hat{w},Q,Q_r,\tilde{Q}) \right] \times \nonumber \\
& \times &
            \f{D_{c}^{\pi} (z_c, \tilde{Q}^2)}{\pi z_c^2}  \,\,  . 
\end{eqnarray}
Here $t_{A}(b)= \int \dd z \, \rho_{A}(b,z)$ denotes the nuclear thickness 
function,  and it is normalized as usual: 
$\int \dd ^2b \, t_{A}(b) = A$. For proton and deuteron we use a sharp 
sphere approximation, while for heavy nuclei the Wood-Saxon formula 
is applied for the nuclear density distribution, $\rho_{A}(b,z)$.

In our next-to-leading order (NLO) calculation, 
$\dd {\widehat \sigma}/ \dd \hat{v}$ represents the Born cross 
section of the partonic subprocess and 
$K_{ab,c}(\hat{s},\hat{v},\hat{w},Q,Q_R,Q_F)$ is the corresponding 
higher order correction term, see Ref.s~\cite{Aversa89,Aur00,pgNLO}. 
We fix the factorization and renormalization scales and connect them 
to the momentum of the intermediate jet, $Q=Q_R=(4/3) p_q$ (where 
$p_q=p_T/z_c$), reproducing $pp$ data with high precision at high 
$p_T$~\cite{dAu}.

To take into account the transverse momentum distribution, we
defined the following $3$-dimensional parton distribution function 
(PDF): 
\begin{equation}
f_{a/p}(x_a,{\bf k}_{Ta},Q^2) \,\,\,\, = \,\,\,\, f_{a/p}(x_a,Q^2) 
\cdot g_{a/p} ({\bf k}_{Ta}) \ .
\end{equation}
Here, the function $f_{a/p}(x_a,Q^2)$ represents the standard 
longitudinal NLO PDF as a function of momentum fraction of the 
incoming parton, $x_a$, at the factorization scale $Q$. 
In the present calculation we use the 
MRST(cg) parameterization~\cite{MRST01}. The partonic 
transverse-momentum distribution in $2$ dimensions, 
$g_{a/p}({\bf k}_T)$, is characterized by an "intrinsic $k_T$" 
parameter as in Refs.~\cite{Yi02,pgNLO}. In our phenomenological 
framework we assumed a Gaussian function~\cite{Yi02,Bp02}.

Nuclear multiscattering is accounted for through a broadening of the 
incoming parton's transverse momentum distribution function, namely 
an increase in the width of the Gaussian:
\beq
\label{ktbroadpA}
\k2av_{pA} = \k2av_{pp} + C \cdot h_{pA}(b) \ .
\eeq
Here, $\k2av_{pp}=2.5$ GeV$^2$ is the width of the transverse 
momentum distribution of partons in $pp$ collisions~\cite{Yi02,dAu}, 
$h_{pA}(b)$ describes the number of {\it effective} $NN$ collisions 
at impact parameter $b$, which impart an average transverse momentum 
squared $C$. The effectivity function $h_{pA}(b)$ can be written in 
terms of the number of collisions suffered by the incoming proton in 
the target nucleus. In Ref.~\cite{Yi02} we have found  a limited 
number of semihard collisions, $3 \leq \nu_{m} \leq 4$ and the value 
$C = 0.35$ GeV$^2$.

We take into account the isospin asymmetry by using a linear 
combination of $p$ and $n$ PDFs. The applied PDFs are also modified 
inside nuclei by the ``shadowing'' effect, applying the
parametrization of Ref.~\cite{Shadxnw_uj}. 

The last term in the convolution of eq. (\ref{hadX}) is the 
fragmentation function (FF), $D_{c}^{\pi}(z_c, \tilde{Q}^2)$. This 
gives the probability for parton $c$ to fragment into a pion with 
momentum fraction $z_c$ at fragmentation scale $\tilde{Q}=(4/3) p_T$. 
We apply the KKP parameterization~\cite{KKP}.

Finally we include jet-quenching as a final state effect.  
The energy loss of high-energy quarks and gluons traveling through  
dense colored matter can measure the integrated density of the colored
particles. This non-Abelian radiative energy loss, $\Delta E (E,L) $, can 
be described as a function of parton opacity, 
(or mean number of jet scatterings), $\bar{n}=L/ \lambda$,
where $L$ is the interaction length of the 
jet and $\lambda$ is the mean free path in non-Abelian dense 
matter. In "thin plasma" approximation the energy loss to first order is 
given by the following expression~\cite{glv}:
\begin{eqnarray}
\label{glv}
\Delta E^{(1)}_{GLV}
&=& \frac{2 C_R \alpha_s}{ \pi} \frac{EL}{\lambda} \int\limits^1_0 \dd x 
    \int\limits^{k_{max}^2}_0 \frac{\dd {\bf k}^2_T}{{\bf k}^2_T} \times 
    \nonumber \\ 
& \times & \int\limits^{q^2_{max}}_0 \frac{\dd ^2 {\bf q}_T \mu^2_{eff}}
    { \pi \left( {\bf q}^2_{T} + \mu^2 \right) ^2} \cdot
    \frac{2 {\bf k}_T \cdot {\bf q}_T \left( {\bf k} -{\bf q} \right) ^2_T L^2}
    {16\,x^2\,E^2 +\left( {\bf k} -{\bf q} \right) ^4_T L^2} \nonumber \\  
&=& \frac{C_R \alpha_s}{N(E)} \frac{L^2 \mu^2}{\lambda} 
\log\left( \frac{E}{\mu} \right) \,\,\, .
\end{eqnarray}
Here $C_R$ denotes the color Casimir of the quark or gluon jet, 
$\mu/\lambda \sim \alpha_s^2 \rho_{part}$ is the transport coefficient 
of the medium, which is proportional to the parton density, $\rho_{part}$, and 
$\mu$ denotes the color Debye screening scale. 
$N(E)$ is an energy dependent factor 
with asymptotic value 4.   

Considering a time-averaged, static plasma, the average energy loss, 
$\Delta E$, will modify the argument of the FFs:
\begin{equation}
\label{quenchff}    
\frac{D_{\pi/c} ( z_c , \tilde{Q}^2 )}{\pi z_c^2 } \longrightarrow 
\frac{z^{\ast}_c}{z_c}  \,\, 
\frac{D_{\pi/c} ( z^{\ast}_c , \tilde{Q}^2 )}{\pi z_c^2 }.
\end{equation}
Here $ z^{\ast}_c = z_c / \left(1- \Delta E/p_c \right) $ is the 
modified momentum fraction.

We will present results on hadron productions 
through the nuclear 
modification factor, $R_{AA'}(p_T,b)$ defined by
\begin{equation}\label{raa_def}
R_{AA'}(p_T,b)=\frac{1}{N_{bin}} \cdot 
\frac{E_{\pi} \dd \sigma^{AA'}_{\pi}(b)/\dd^3 p}
{E_{\pi} \dd \sigma^{pp}_{\pi}/\dd^3 p} \, \, .
\end{equation}
Here $N_{bin}$ is the average number of binary collisions,
as a function of $p_T$ at different impact parameter ranges.

\section{Jet Tomography in Different $AA$ Systems}
\label{sec:2}

The opacity parameter, $\bar{n}=L/\lambda$, can be determined by
finding the best fit for energy loss and comparing the theoretical
results to the data points on the nuclear modification factor.
In this section we present  our
jet tomography results in the mid-rapidity region.  

\begin{figure}
\resizebox{0.5\textwidth}{!}{%
  \includegraphics{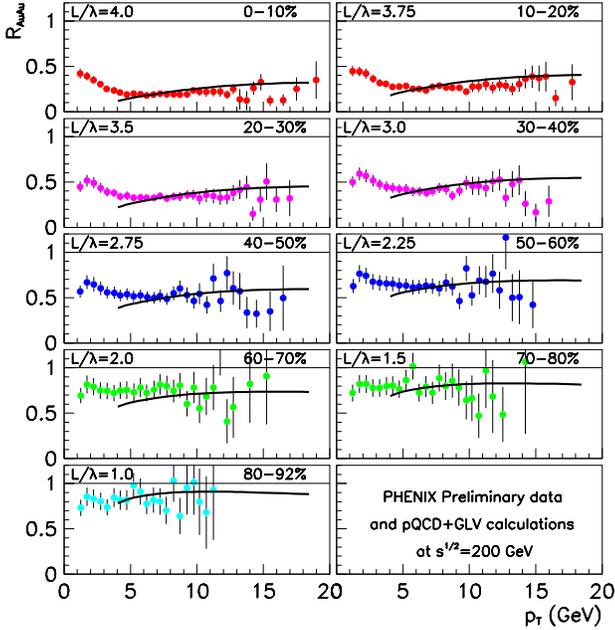}
}
\caption{The nuclear modification factor for $\pi^0$ production
in $AuAu$ collision, 
$R^{\pi}_{AuAu}(p_T)$, at energy $\sqrt{s}=200~A$GeV at mid-rapidity,
measured by PHENIX~\cite{PHENIXdau05,PHENIXHQ06}. The solid lines indicate our 
best fit with the applied opacity parameters, $L/\lambda$.}
\label{fig:1}      
\end{figure}

Fig.~\ref{fig:1} displays our fit on opacity in $AuAu$ collision
at different centralities, 
using the preliminary data of the PHENIX collaboration on 
$\pi^0$ production at mid-rapidity at 
$\sqrt{s}=200$ $A$GeV energy~\cite{PHENIXdau05,PHENIXHQ06}.  
Experimental data indicate a stronger suppression pattern at
$p_T> 4$ GeV/c, thus we read the $L/\lambda$ values 
for this region. Here the nuclear modification factor is flat,
as Fig.~\ref{fig:1} displays.
The extracted opacity values are boxed 
at the top of the panels, together with the centralities. The  plotted
$R^{\pi}_{AuAu}(p_T)$ are calculated at the mean values of the 
opacity, the errors of the fits are approximately $\pm 0.25$.  

The measured suppression pattern in $AuAu$ collision is reproduced well
in all centrality bins. In the most central ($0-10\%$) collisions
we have obtained a maximal opacity value, $L/\lambda=4.0 \pm 0.25$. 
In the most peripheral collisions ($80-92\%$)
we have extracted the opacity $L/\lambda= 1.0 \pm 0.25$. 

\begin{figure}
\resizebox{0.5\textwidth}{!}{%
  \includegraphics{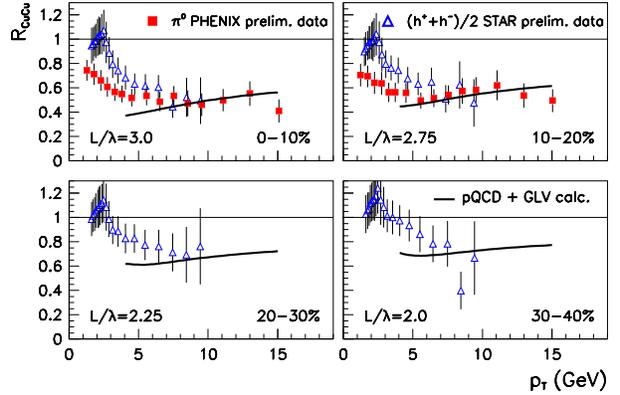}
}
\caption{The nuclear modification factor for pion production
in $CuCu$ collision, 
$R^{\pi}_{CuCu}(p_T)$, in mid-rapidity at energy $\sqrt{s}=200~A$GeV.
Data are from PHENIX ($\pi^0$)~\cite{PHENIXCuCu} and STAR
($\frac{h^+ + h^-}{2}$)~\cite{STARCuCu}. The solid lines indicate our 
best fit with the applied opacity parameters, $L/\lambda$.}
\label{fig:2}      
\end{figure}

We analyzed  the data from $CuCu$ collisions at 
$\sqrt{s}=200~A$GeV energy in a similar fashion. Fig.~\ref{fig:2} displays
the existing data and our calculations. We have found preliminary 
data on $\pi^0$ production from the PHENIX Collaboration
in the $0-10 \%$ and  
$10-20 \%$ centrality bins~\cite{PHENIXCuCu}.
We are able to calculate pion spectra with high precision in our pQCD 
framework, thus we compare data and theory directly in these
two centrality bins.
In parallel, STAR data exist on unidentified charged hadrons
($h^{\pm}:=\frac{h^++h^-}{2}$) for more centrality 
bins~\cite{STARCuCu}.
In the most central bins we see that the nuclear modification
factors overlap for identified pions and unidentified charge 
hadrons in the very high-$p_T$ region, $p_T> 6-8 $ GeV/c.
(This indicates the restoration of the fragmentation limit for 
proton and antiproton production in this region,
thus the validity of pQCD based calculation for all hadron species
--- for smaller momenta the contribution from quark coalescence
should be included for baryon and antibaryon 
production~\cite{COA1,COA2}).
Thus, focusing on the highest momentum region, we extracted the
opacity values from the STAR data on unidentified charge hadrons
and display the data and the theoretical results on 
Fig.~\ref{fig:2} for two more centrality bins,
namely for $20-30\% $ and $30-40\% $.

\begin{figure}
\resizebox{0.5\textwidth}{!}{%
  \includegraphics{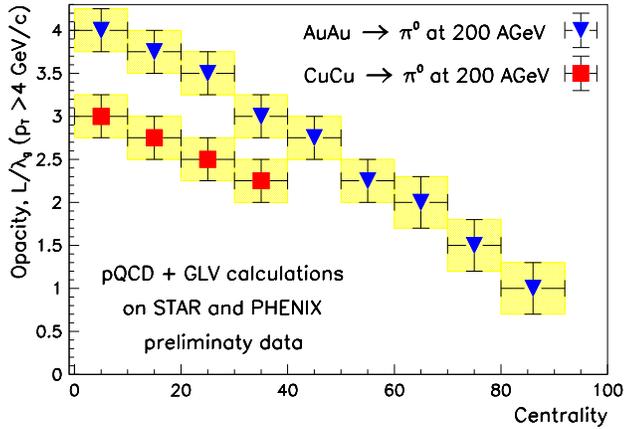}
}
\caption{The centrality dependence of the opacity parameter, $L/\lambda$, 
in $AuAu$ ({\it triangles}) and $CuCu$ ({\it squares})
collisions at energy $\sqrt{s}=200~A$GeV.}
\label{fig:3}      
\end{figure}

Fig.~\ref{fig:3} summarizes all mid-rapidity results
on  opacity  for $AuAu$ collisions in the $0-92\%$
centrality region ({\it triangles}) and for $CuCu$ collisions in 
the $0-40\%$ region ({\it squares}).
The opacity parameter, which is connected to the color 
charge density, is decreasing approximately linearly with increasing  
centrality, and the slopes of the opacity curves are very
similar for $AuAu$ and $CuCu$ collisions. Furthermore,
the mid-central ($30-40\%$) $AuAu$ 
collisions are characterized by the same opacity parameter 
as the most central ($0-10\%$) $CuCu$ case. 
This overlap can be seen clearly in the measured nuclear
modification factors for $AuAu$ and $CuCu$ collisions
 at different centralities and it 
has been discussed
in Refs.~\cite{phenixauau,Henner_talk,PHOBOS06}.
This effect may present a real challenge to theoretical 
models as well as the consequences of any explanation.

\begin{figure}
\resizebox{0.5\textwidth}{!}{%
  \includegraphics{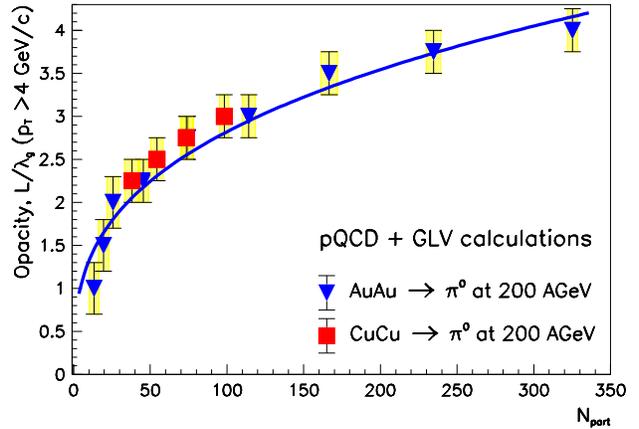}
}
\caption{The opacity parameter, $L/\lambda$, as a function
of the participant number, 
$\langle N_{part} \rangle $ in $AuAu$ ({\it triangles})
and $CuCu$ ({\it squares}) collisions at $\sqrt{s}=200~A$GeV.}
\label{fig:4}      
\end{figure}

Fig.~\ref{fig:4} displays the extracted opacity parameters as a function
of the participant number, $\langle N_{part} \rangle$, in $AuAu$
collisions ({\it triangles}) and $CuCu$ collisions ({\it squares}). 
Here the overlap of the two centrality dependent
opacity parameter sets is clearly seen.  
Fitting the opacity values with a continuous function
of the participant number (see solid line on Fig.~\ref{fig:4}), 
the following functional form has been found: 
\begin{equation} 
L/\lambda = (0.62 \pm 0.09) \cdot \langle N_{part}\rangle ^{0.33 \pm 0.03} \, .
\end{equation}
This fit agrees well with the simple expectation based on geometry, namely
$L \propto A^{1/3} \propto N_{part}^{1/3}$~\cite{IvanQM05}.

Assuming constant mean free path ($\lambda$) and zero formation time,
the above fit validates the idea of a volume-dominated jet energy loss 
for the whole volume of the hot dense matter. This was assumed
in the GLV-method~\cite{glv}, thus our analysis is self-consistent. 
Combining this result with the expression of the induced energy loss
in eq.~(\ref{glv}), one can obtain the simple connection,
$\epsilon = \Delta E/E \propto N_{part}^{2/3}$.
In the forthcoming sessions we will assume the presence of
volume-dominated energy losses. 

On the other hand, in models which lead to surface-dominated 
jet-energy loss~\cite{eskola,guy,Pantuev} or which assume 
non-zero formation times~\cite{peigne}  
the above picture is modified drastically. It would be interesting
to see the centrality dependence in a repeated analysis with
these assumptions and the appropriate plots equivalent 
to Fig.~\ref{fig:3} and Fig.~\ref{fig:4}. However, this is not
our aim, here, in this paper.

\section{Jet Tomography at Large Rapidities}

The jet-quenching pattern of the nuclear modification factor is driven 
by the space and time averaged value of the color-charge density.  
We are interested in the presence of possible asymmetry 
between transverse and longitudinal directions seen by the particles
produced in different rapidities.
Since the produced hot dense matter is characterized by different dynamical
behavior in transverse and longitudinal direction, we expect to
see this difference during the analysis of available data. 
Here we analyze hadron production in the forward rapidity
region using our theoretical framework and the latest data at RHIC.

In the latest BRAHMS results
obtained in the most central $AuAu$ collisions~\cite{BRAHMS05,BRAHMS}
the nuclear modification factor,
$R_{AA}(p_T)$, displays similar values at different
rapidities, namely at $y=0$ and $y=3.2$.
We expect, that the combination of appropriate geometry of the
hot dense partonic region and the competing nuclear effects, namely
shadowing, multiscattering, and induced energy loss explain this
result. Thus, if we extend our pQCD based calculation and include these
latter nuclear effects (which seems to be under control
at higher rapidities~\cite{Barnafqm04,bggqm05}), 
then the analysis of high rapidity data
offers a unique possibility to extract information about the 
properties of the dense matter in the longitudinal direction
in the very early stage of the heavy ion collisions.

The BRAHMS data~\cite{BRAHMS05,BRAHMS} are available in the 
inter\-mediate-$p_T$ region ($p_T < 4$~GeV/c). We will assume
that the suppression factors at different rapidities
are approximately the same in the high-$p_T$ region
and we will perform our calculations
in this manner. This means a suppression factor $\sim 5$ 
for $AuAu$ collisions at $\sqrt{s}=200$~$A$GeV (seen in the
most central collisions in the mid-rapidity~\cite{PHENIX_QM05})
and a suppression factor $\sim 3$  
at $\sqrt{s}=62.4$~$A$GeV (see PHENIX data in the most central 
$AuAu$ collisions~\cite{PHENIXHQ06}.) 

\newpage

\begin{figure}
\resizebox{0.5\textwidth}{!}{%
  \includegraphics{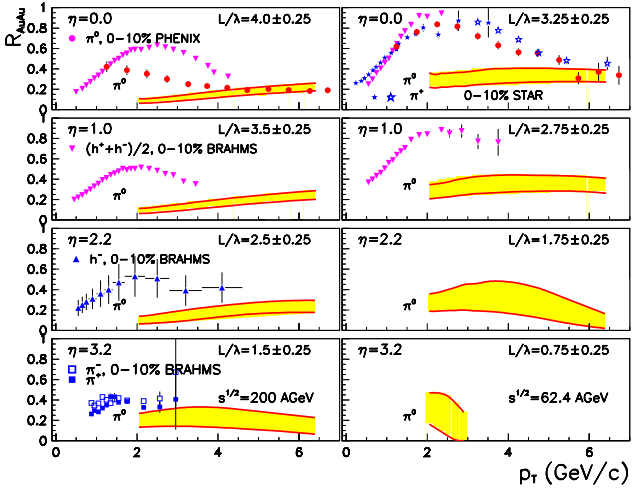}
}
\caption{The pQCD results and the appropriate
opacities, $L/\lambda$, at different rapidities 
for pion production in $AuAu$ collisions at $\sqrt{s}=200~A$GeV 
({\it left panels}) and $62.4$~$A$GeV ({\it right panels}). 
Data are from BRAHMS~\cite{BRAHMS05,BRAHMS}, 
PHENIX~\cite{PHENIX_QM05,PHENIXHQ06},
STAR~\cite{STARCuCu}.}
\label{fig:5}      
\end{figure}

Fig.~\ref{fig:5} displays our results in the (pseudo)rapidity regions
$\eta=0.0$, $1.0$, $2.2$ and $3.2$ for central $AuAu$ collisions 
at energies  $\sqrt{s}=200$~$A$GeV ({\it left panels}) and  
$62.4$~$A$GeV ({\it right panels}). Data are from 
BRAHMS~\cite{BRAHMS05,BRAHMS} at higher rapidities.
The mid-rapidity data are from  
PHENIX~\cite{PHENIX_QM05,PHENIXHQ06} on pion and from
STAR~\cite{STARCuCu} on charged hadron production.
We display the extracted $L/\lambda$ opacity values for all 
rapidities. The extended bands 
indicate the theoretical uncertainties on the induced energy 
loss denoted by the errors in the opacity values. 

\begin{figure}
\resizebox{0.522\textwidth}{!}{%
  \includegraphics{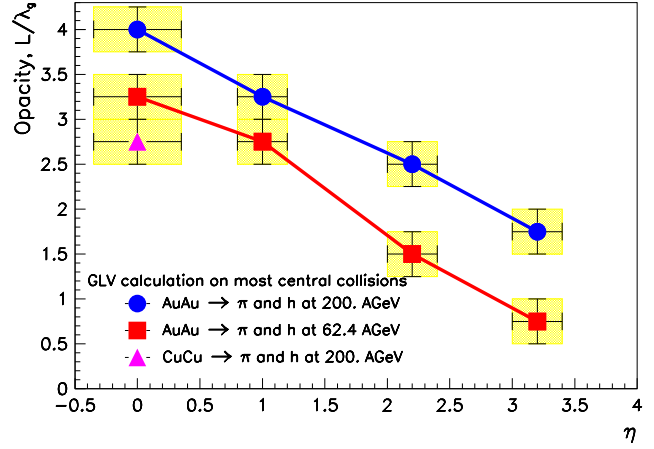}
}
\caption{The opacity parameter $L/\lambda$ as the function of  
pseudorapidity ($\eta$) for forward pion production in the
most central $AuAu$ and $CuCu$ collisions at energies
$\sqrt{s} = 62.4, \ 200$~$A$GeV, and 
$CuCu$ collisions at energies $\sqrt{s} = 200$~$A$GeV.}
\label{fig:6}      
\end{figure}

Fig.~\ref{fig:6} summarizes the obtained opacity parameters in wide
rapidity region for the most central ($0-10\%$) $AuAu$ 
collisions at  $\sqrt{s}=200~A$GeV and $62.4$~$A$GeV energies.
(We have included the opacity parameter from  $CuCu$ collisions
at $\sqrt{s}=200~A$GeV.) 
In the highest energy domain 
the opacity parameter is decreasing with increasing rapidity from
the maximal value of $L/\lambda= 4.0\pm 0.25$ at mid-rapidity
to $L/\lambda= 1.5\pm 0.25$ at $\eta=3.2$. 
Decreasing the collision energy to $\sqrt{s} = 62.4$~$A$GeV 
a smaller opacity, $L/\lambda= 3.25\pm 0.25$ has been extracted
at $\eta=1$. Since no data available at higher rapidities at this energy,
we assumed the unmodified suppression factor. This way we have obtained
$L/\lambda= 0.75\pm 0.25$ at $\eta=3.2$. Fig.~\ref{fig:5} can indicate
the validity of this assumption.
Fig.~\ref{fig:6} clearly shows an
approximately linear decreasing tendency for the opacity parameter as 
$\eta$ is increasing. We have found that at large forward rapidities the 
interplay between a stronger shadowing and a linearly (in $\eta$)
decreasing quenching effect is able to 
maintain a rapidity independent nuclear modification factor. 

Fig.~\ref{fig:6} displays one more interesting message, 
what we were looking for.
Namely, comparing the extracted opacity values in the mid-rapidity and 
in the most forward rapidity, one can see a factor of 3 difference
at both energies, $\sqrt{s} = 62.4$~$A$GeV and $\sqrt{s} = 200$~$A$GeV.
According to our assumption the produced hot dense deconfined matter
is homogeneous, thus we have a uniform $\lambda$ value (which may
depend on energy but not on geometry). 
Thus the above result indicates that
longitudinally traveling partons see less 
colored matter than those traveling in the transverse direction.
This result can be explained on the basis of pure geometry
combined with dynamics.

\begin{figure}[b]
\resizebox{0.5\textwidth}{!}{%
  \includegraphics{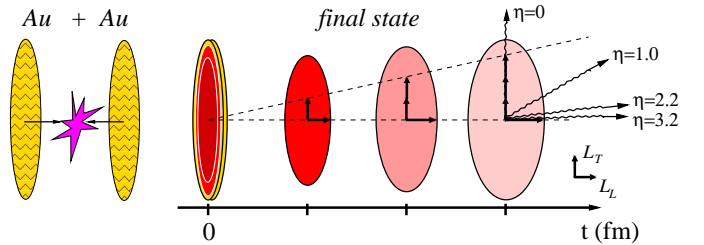} 
}
\caption{Time evolution of the quark or gluon jet traveling through 
the expanding hot dense matter into the transverse and longitudinal
direction.}
\label{fig:8}      
\end{figure}

Fig.~\ref{fig:8} displays a schematic picture for the
time evolution of the formed hot dense matter and the outcoming jets.
Let us characterized the hot matter with the mean free path of
$\lambda=1$ fm.
A jet, created in the central region of the collision and producing
mostly mid-rapidity hadrons, is traveling transversally through an 
$L_T \sim 4$ fm length. This jet looses a large portion of its energy and
indicates an opacity $L_T/\lambda=4$.
Hadrons in the forward rapidities are produced from forward jets. 
These jets are moving mostly longitudinally and after passing
a (contracted) thin region of the compressed matter 
(characterized by an effective length of $L_L \sim 1.5$ fm) 
they reach very quickly the longitudinally expanding surface.
Both jet and expansion surface are moving with speed of light,
thus the comoving jet does not loose more energy and 
an opacity $L_L/\lambda = 1.5$ can be extracted at the 
highest rapidities.

\newpage

Decreasing the collision energy, the density of the produced hot matter
is decreasing and the mean free path is increasing. We do not expect
strong modification in the geometry and the comoving dynamics.
Thus our results at 
$\sqrt{s} = 62.4$~$A$GeV indicate a 25 \% increase in the mean free path
comparing to the case of $\sqrt{s} = 200$~$A$GeV. 


\section{Summary}

We have analyzed the measured suppression pattern in
$AuAu$ and $CuCu$ collisions
at energy $\sqrt{s}=200$ $A$GeV at mid-rapidity. We have found the
opacity of the strongly interacting hot matter to be proportional
to the participant nucleon number. This result indicates
the presence of drastical microscopical mechanisms for energy stopping,
which are working in a uniform way at given energy and producing similar
energy densities in heavy ion collisions of different nuclei.

Investigating hadron production in forward rapidities,
the suppression pattern indicated the formation of a longitudinally
contracted dense deconfined zone in the most central heavy ion collisions.
We could have determined the initial geometry from the existing data.
The decrease of the opacity into the longitudinal direction
is explained by the appearance of comoving dynamics.

Our one-hadron tomography can be combined with
di-hadron correlation studies~\cite{PHENIXdih,LFP06} to obtain 
more detailed information  about the geometry 
of the hot region. Energy and centrality dependence in $CuCu$
collisions could be used to verify our results.

\vspace*{-0.2truecm}

\section*{Acknowledgments}

One of the authors (GGB) would like thank the support of the
Organizers. This work is also supported by Hungarian OTKA T043455, 
T047050, NK62044, MTA-NSF-OTKA OISE-0435701 
and U.S. DE-FG02-86ER40251 grants.
%

\begin{thebibliography}{99}
%

\bibitem{PHENIX_QM01}
 G. David {\it et al.} (PHENIX Coll.), 
    Nucl. Phys. {\bf A698} (2002) 227.

\bibitem{PHENIX_QM05}
 V. Greene {\it et al.} (PHENIX Coll.), 
    Nucl. Phys. {\bf A774} (2006) 93. 

\bibitem{STAR_QM05}
    J. Dunlop {\it et al.} (STAR Coll.), 
    Nucl. Phys. {\bf A774} (2006) 139. 

\bibitem{Levai_QM01}
    P. L\'evai {\it et al.}
     Nucl. Phys. {\bf A698} (2002) 631.

\bibitem{GV02}
    I. Vitev and M. Gyulassy,
    Phys. Rev. Lett. {\bf 89} (2002) 252301.

\bibitem{glv}
    M. Gyulassy, P. L\'evai, I. Vitev,
    Phys. Rev. Lett. {\bf 85} (2000) 5535; 
    Nucl. Phys. {\bf B571} (2000) 197; 
    {\bf B594} (2001) 371.

\bibitem{zakhar}
    B.G. Zakharov, JETP Lett. {\bf 63} (1996)  952; {\bf 64} 
    (1996) 781; {\bf 65} (1997) 615; {\bf 70} (1999) 176;  {\bf 73} (2001) 
    49; {\bf 80} (2004) 617. Phys. Atom. Nucl.{\bf  61} (1998) 838.


\bibitem{bdms}
    R. Baier, Yu.L. Dokshitzer, A.H. Mueller, S. Peigne, D. Schiff,
    Nucl. Phys. {\bf B483} (1997) 291; {\bf B484} (1997) 265;
    {\bf B531} (1998) 403.

\bibitem{urs}
    U.A. Wiedemann, Nucl. Phys. {\bf B582} (2000) 409;
    Nucl. Phys. {\bf B588} (2000) 303;
Nucl. Phys. {\bf A690} (2001) 731.


\bibitem{bgg04}  
    G.G. Barnaf\"oldi {\it et al.},
    {Eur. Phys. J.} {\bf C33} (2004) S609.

\bibitem{PHENIX03}
    D. d'Enterria {\it et al.} (PHENIX Coll.),
    Nucl. Phys. {\bf A715} (2003) 749.

\bibitem{PHENIX03a}
    S.S. Adler {\it et al.} (PHENIX Coll.),
    Phys. Rev. Lett. {\bf 91} (2003) 072301.

\bibitem{BRAHMS05}
    P. Staszel {\it et al.} (BRAHMS Coll.), 
    Nucl. Phys. {\bf A774} (2006) 77.

\bibitem{BRAHMS}
    D. R\"ohrich {\it et al.} (BRAHMS Coll.), 
    Nucl. Phys. {\bf A774} (2006) 297.

\bibitem{BRAHMSdau04}
   I. Arsene {\it et al.} (BRAHMS Coll.),
    {Phy. Rev. Lett.} {\bf 93} (2004) 242303; 
     Z. Yin {\it et al.} (BRAHMS Coll.), 
    {J. Phys.} {\bf G30} (2004) S983;
     Z. Yin for the BRAHMS Coll.,
     Acta Phys Hung. {\bf A22} (2005) 309.

\bibitem{Barnafqm04}  
    G.G. Barnaf\"oldi {\it et al.}, 
    {J. Phys.} {\bf G30} (2004) S1125.

\bibitem{bggqm05}
    G.G. Barnaf\"oldi {\it et al.}, 
    Nucl. Phys. {\bf A774} (2006) 801.

\bibitem{Yi02}
    Y. Zhang {\it et al.}, 
    {\it Phys. Rev.} {\bf C65} (2002) 034903.

\bibitem{dAu}
    P. L\'evai {\it et al.},
    {\tt nucl-th/0306019}.

\bibitem{Colum05}
    M. Djordevic, M. Gyulassy, S. Wicks,
    Eur. Phys. J. {\bf C43} (2005) 135.

\bibitem{Colum06}
    A.Adil, M. Gyulassy, W.A. Horowitz, S. Wicks,
    {\tt nucl-th/0606010}.

\bibitem{Aversa89}
F. Aversa, P. Chiappetta, M. Greco, and J.Ph. Guillet, \\
Nucl. Phys. {\bf B327} (1989) 105.

\bibitem{Aur00}
P. Aurenche, M. Fontannaz, J.Ph. Guillet, B. Kniehl, E. Pilon, and M. Werlen,
Eur. Phys. J. C {\bf 9} (1999) 107; \\
P. Aurenche, M. Fontannaz, J.Ph. Guillet, B. Kniehl, and M. Werlen,
Eur. Phys. J. C {\bf 13} (2001) 347.

\bibitem{pgNLO}
    G. Papp {\it et al.},
    {\tt hep-ph/0212249}.

\bibitem{MRST01}  
    A.D. Martin {\it et al.},
    {\it Eur. Phys. Jour.} {\bf C23} (2003) 73.

\bibitem{Bp02}  
    G.G. Barnaf\"oldi {\it et al.},
    {Heavy Ion Phys.} {\bf 18} (2003) 79.

\bibitem{Shadxnw_uj}
    S.J. Li and X.N. Wang, 
    {Phys. Lett.} {\bf B527} (2002) 85.

\bibitem{KKP}
    B.A. Kniehl, G. Kramer, and B. P{\"o}tter, 
    {Nucl. Phys.} {\bf B597} (2001) 337.

\bibitem{PHENIXdau05}
    B.A. Cole, 
    Nucl. Phys. {\bf A774} (2006) 225.

\bibitem{PHENIXHQ06}
    H. B\"usching for the PHENIX Coll.,
    Talk on the Hot Quark'06 Conference.
    See this Proceedings.

\bibitem{PHENIXCuCu}
    C. Klein-B\"osing for the PHENIX Coll.,
    Proceedings of the 22nd Winter Workshop on Nuclear
    Dynamics, La Jolla, California, USA, 2006.
    To appear in Acta Phys. Hung. {\bf A}
    ({\tt nucl-ex/0606013}).
    
\bibitem{STARCuCu}
     J. Dunlop for the STAR Coll.,
     Talk on the Quark Matter'05 Conference,
     2005, Budapest, Hungary.
    
\bibitem{COA1}
        V. Greco, C.M. Ko, P. L\'evai,
        Phys. Rev. Lett. {\bf 90}, 202302 (2003);
        Phys. Rev. {\bf C68}, 034904 (2003).

\bibitem{COA2}
        R.J. Fries, B. M\"uller, C. Nonaka, S. Bass,
        Phys. Rev. Lett. {\bf 90}, 202303 (2003);
        Phys. Rev. {\bf C68}, 044902 (2003).

\bibitem{phenixauau}
   S.S. Adler {\it et al.} (PHENIX Coll.),
    {Phy. Rev. Lett.} {\bf 91} (2003) 072301; 

\bibitem{Henner_talk}
   H. B\"usching  {\it et al.} (PHENIX Coll.),
    Nucl. Phys. {\bf A774} (2006) 103.

\bibitem{PHOBOS06}
   B. Alver {\it et al.} (PHOBOS Coll.),
    {Phy. Rev. Lett.} {\bf 96} (2006) 212301; 

\bibitem{IvanQM05}
        I. Vitev, {\tt hep-ph/0511237},
        To appear in Acta Phys. Hung. (2006).

\bibitem{eskola}
    K.J. Eskola, H. Honkanen, C.A. Salgado, U.A. Wiedemann,
    Nucl. Phys. {\bf A747} (2005) 511.

\bibitem{guy}
    A. Dainese, C. Loizides, G. Paic,
    Eur. Phys. J. {\bf C38} (2005) 461.

\bibitem{Pantuev}
    V. Pantuev,
    {\tt hep-ph/0506095}.

\bibitem{peigne}
    S. Peigne, P.-B. Gossiaux, T. Gousset,
    JHEP 0604 (2006) 011.

\bibitem{PHENIXdih}
    S.S. Adler {\it et al.} (PHENIX Coll.),
    Phys. Rev. Lett. {\bf 96} (2006) 222301;
     {\tt hep-ex/0605039}.

\bibitem{LFP06}
    P. L\'evai, G. Fai, G. Papp,
    Phys. Lett. {\bf B634} (2006) 383.

\end{thebibliography}
\vspace*{-0.2truecm}
%

\end{document}